\documentclass[prd,showpacs]{revtex4}
\usepackage{amsmath,amssymb}
\DeclareMathOperator{\csch}{csch}
\begin{document}
\title{Casimir effect of a Lorentz-violating scalar in magnetic field}
\author{Andrea Erdas}
\email{aerdas@loyola.edu}
\affiliation{Department of Physics, Loyola University Maryland, 4501 North Charles Street,
Baltimore, Maryland 21210, USA}
\begin {abstract} 
IIn this paper I study the Casimir effect caused by a charged and massive scalar field that breaks Lorentz invariance in a CPT-even, aether-like manner. I investigate the case of a scalar field that satisfies Dirichlet or mixed (Dirichlet-Neumann) boundary conditions on a pair of very large plane parallel plates. The case of Neumann boundary conditions is straightforward and will not be examined in detail. I use the $\zeta$-function regularization technique to study the effect of a constant magnetic field, orthogonal to the plates, on the Casimir energy and pressure. I investigate the cases of a timelike Lorentz asymmetry, a spacelike Lorentz asymmetry in the direction perpendicular to the plates, and a spacelike asymmetry in the plane of the plates and, in all those cases, derive simple analytic expressions for the zeta function, Casimir energy and pressure in the limits of small plate distance, strong magnetic field and large scalar field mass. I discover that the Casimir energy and pressure, and their magnetic corrections, all strongly depend on the direction of the unit vector that implements the breaking of the Lorentz symmetry.
\end {abstract}
\maketitle
\section{Introduction}
\label{1}
The first theoretical prediction of an attractive force of a purely quantum nature between two uncharged and conducting parallel plates in vacuum is due to Hendrik Casimir  \cite{Casimir:1948dh}. Ten years later sophisticated experiments, for that time, were done by Sparnaay \cite{Sparnaay:1958wg} who tested Casimir's prediction and obtained results that were not inconsistent with it, even though were not accurate enough to confirm it. The Casimir effect was later confirmed by many and increasingly accurate experimental verifications  \cite{Bordag:2001qi,Bordag:2009zz} that followed Sparnaay's. The Casimir effect strongly depends on the boundary conditions at the plates of the quantum field under consideration. Dirichlet or Neumann boundary conditions cause an attraction between the plates, mixed (Dirichlet-Neumann) boundary conditions cause a repulsive force \cite{Boyer:1974}. 

While in standard quantum field theory Lorentz invariance is strictly obeyed, other theories propose models where a violation of Lorentz symmetry leads to a space-time anisotropy \cite{Ferrari:2010dj,Ulion:2015kjx}. Mechanisms for breaking Lorentz symmetry have been proposed in modified quantum gravity \cite{Alfaro:1999wd,Alfaro:2001rb}, in theories that propose variation of some coupling constants \cite{Kostelecky:2002ca,Anchordoqui:2003ij,Bertolami:1997iy}, and in string theory \cite{Kostelecky:1988zi}, where non-vanishing expectation values of some vector and tensor fields components lead to a spontaneous violation of Lorentz invariance at the Planck energy scale. A comprehensive list of papers that study various consequences of Lorentz symmetry breaking is available in the work by Cruz et al. \cite{Cruz:2017kfo}. Implications of a Lorentz asymmetry in the Casimir effect have been studied in the case of Lorentz-breaking extensions of QED \cite{Frank:2006ww,Kharlanov:2009pv,Martin-Ruiz:2016lyy}, and in the case of a real scalar field  in vacuum \cite{Cruz:2017kfo} and in a medium at finite temperature \cite{Cruz:2018bqt}. Both of these papers investigate a modified Klein-Gordon model that breaks Lorentz symmetry in a CPT-even, aether-like manner. 

While several authors have studied magnetic corrections to the Casimir effect in a symmetric spacetime with unbroken Lorentz symmetry \cite{CougoPinto:1998td,CougoPinto:1998jg,Erdas:2013jga}, there has not been a study of magnetic corrections to the Casimir effect of a charged scalar field that breaks the Lorentz symmetry. This work intends to fill that gap and provide theoretical predictions of the magnetic field effects on the quantum vacuum of the modified Klein-Gordon model introduced in Ref. \cite{Cruz:2017kfo}.
In this paper I will investigate the effect of a uniform magnetic field  $\vec B$ on the Casimir energy and pressure due to a Lorentz-violating scalar field, by studying a model similar to the one first presented in Ref.  \cite{Cruz:2017kfo}: a charged scalar field that breaks Lorentz symmetry and satisfies either Dirichlet or mixed boundary conditions on a pair of very large parallel plates at distance $a$ from each other.

The Casimir force between rigid objects is always finite, as the TGTG formula \cite{Kenneth:2006vr} shows, but direct calculations of the Casimir energy often require a regularization method. In the case of parallel plates a naive calculation of the energy of the vacuum between the plates, without regularization, will produce an infinite result. While Ref. \cite{Cruz:2017kfo} uses the Abel-Plana method to regularize the Casimir energy, in this study I will use the zeta function regularization technique \cite{Elizalde:1988rh,Elizalde:2007du}, a modern method also used extensively within the framework of finite temperature field theory \cite{Santos:1998vb,CougoPinto:1998xn,Santos:1999yj}.

In Sec. \ref{2} of this paper I briefly describe the theoretical model of a charged scalar field that breaks Lorentz symmetry in an aether-like and CPT-even manner, and calculate its zeta function. In Sec. \ref{3} I examine the case of a timelike 
Lorentz asymmetry and calculate the Casimir energy, obtaining simple analytic expressions for the energy in the short plate distance limit, large magnetic field limit, and large mass limit. In Sec. \ref{4} I investigate the case of a spacelike Lorentz 
anisotropy in the same plane as the plates, calculate the Casimir energy, and obtain simple expressions for it in the three limits listed above. In Sec. \ref{5} I focus on a spacelike Lorentz anisotropy perpendicular to the plates, calculate the Casimir energy, and obtain simple expressions for it in the aforementioned three limits. In Sec. \ref{6} I calculate the Casimir pressure for all the cases described above. My conclusions, along with a discussion of my results are presented in Sec. \ref{7}.
\section{The model and its zeta function}
\label{2}

I will investigate the Casimir effect due to a charged scalar field of mass $m$ that breaks the Lorentz symmetry, using the theoretical model of a scalar field that produces an aether-like and CPT-even Lorentz symmetry breaking presented in Ref. \cite{Cruz:2017kfo}. In this model, the Lorentz symmetry violation is caused by a coupling of the derivative of the scalar field to a fixed unit four-vector $u^\mu$. The Klein Gordon equation for this field is
\begin{equation}
[\Box 
+\lambda(u\cdot\partial)^2+m^2]\phi=0,
\label{KG}
\end{equation}
where the dimensionless parameter $\lambda \ll 1$, and the unit four-vector $u^\mu$ indicates the space-time direction in which the Lorentz symmetry is broken. My goal is to study how the space-time anisotropy and the presence of a magnetic field modify the Casimir effect. I am considering two square plates of side $L$ perpendicular to the $z$ axis, and a constant magnetic field $\vec B$ pointing in the $z$ direction. The two plates are located at $z=0$ and $z=a$. I will use the generalized zeta function technique to study this problem, and will investigate Dirichlet and mixed boundary conditions at the plates. The case of Neumann boundary conditions is straightforward and will produce the same results obtained with Dirichlet boundary conditions. I will study the cases when the unit four-vector $u^\mu$ is timelike, spacelike and perpendicular to $\vec B$, and spacelike and parallel to $\vec B$.

I use Euclidean time $\tau$ and begin by writing the eigenvalues of the Klein Gordon operator $D_E$ of Eq. (\ref{KG})
\begin{equation}
D_E=-{\partial^2\over \partial \tau^2}-\nabla^2
+\lambda(u\cdot\partial)^2+m^2,
\label{KGoperator}
\end{equation}
notice that at this initial stage the magnetic field is not present and I will introduce it later.
When $u^\mu$ is timelike and Dirichlet boundary conditions are imposed at the plates, the eigenvalues of $D_E$ are:
\begin{equation}
\left\{(1+\lambda)k_0^2+k_x^2+k_y^2+\left({n\pi\over a}\right)^2+m^2\right\},
\label{eigenvalues1}
\end{equation}
where $k_0, k_x, k_y \in \Re$ and $n=0,1,2,,\cdots$; when one considers mixed boundary conditions, Dirichlet on the plate at $z=0$ and Neumann on the plate at $z=a$, the eigenvalues
of $D_E$ are:
\begin{equation}
\left\{(1+\lambda)k_0^2+k_x^2+k_y^2+\left(n+{1\over 2}\right)^2\left({\pi\over a}\right)^2+m^2\right\}.
\label{eigenvalues1m}
\end{equation}
When $u^\mu$ is spacelike and in the $x$-$y$ plane, I take it as $u^\mu= \left(0,{1\over \sqrt{2}},{1\over \sqrt{2}},0\right)$ and, for Dirichlet boundary conditions, the eigenvalues of $D_E$ are:
\begin{equation}
\left\{k_0^2+\left(1-{\lambda\over 2}\right)(k_x^2+k_y^2)+\left({n\pi\over a}\right)^2+m^2\right\},
\label{eigenvalues2}
\end{equation}
while, for mixed boundary conditions, the eigenvalues of $D_E$ are similar to those listed in Eq. (\ref{eigenvalues2}) with $n$ replaced by $n+{1\over 2}$.
When $u^\mu$ is in the $z$ direction, I find the following eigenvalues of $D_E$ for Dirichlet boundary conditions:
\begin{equation}
\left\{k_0^2+k_x^2+k_y^2+(1-\lambda)\left({n\pi\over a}\right)^2+m^2\right\},
\label{eigenvalues3}
\end{equation}
and the eigenvalues for the case of mixed boundary conditions are obtained replacing $n$ with $n+{1\over 2}$. 

Now I introduce the magnetic field, and its presence modifies the eigenvalues of Eqs. (\ref{eigenvalues1}, \ref{eigenvalues1m}, \ref{eigenvalues2}, \ref{eigenvalues3}) by changing $k^2_x+k^2_y$ into $2eB(\ell+{1\over 2})$, where $e$ is the charge of the scalar field and $\ell=0,1,2,\cdots$, labels the Landau levels. At this point,
I can write the generalized zeta function, with the magnetic field contribution, for the three aforementioned cases of $u^\mu$.
When $u^\mu$ is timelike and Dirichlet boundary conditions are imposed,
\begin{equation}
\zeta(s)=\mu^{2s}{L^2eB\over 2\pi}(1+\lambda)\sum_{n=0}^\infty \sum_{\ell=0}^\infty\int^\infty_{-\infty} {dk_0\over 2\pi} \left[(1+\lambda)k_0^2+(2\ell+1)eB+\left({n\pi\over a}\right)^2+m^2\right]^{-s},
\label{zeta1}
\end{equation}
where the parameter $\mu$ with dimension of mass keeps $\zeta(s)$ dimensionless for all values of $s$, ${L^2eB\over 2\pi}$ takes into account the degeneracy of the Landau levels, and the factor of $(1+\lambda)$ is introduced following Ref. \cite{Cruz:2017kfo} where, in the case of timelike asymmetry, it is shown that the vacuum energy, obtained by taking the vacuum expectation value of the Hamiltonian, carries a factor of $(1+\lambda)$.
For mixed boundary conditions the zeta function is obtained by taking the expression of Eq. (\ref{zeta1}) and replacing $n$ with $n+{1\over 2}$.
When $u^\mu$ is spacelike and perpendicular to $\vec B$, and Dirichlet boundary conditions are considered
\begin{equation}
\zeta(s)=\mu^{2s}{L^2eB\over 2\pi}\sum_{n=0}^\infty \sum_{\ell=0}^\infty\int^\infty_{-\infty} {dk_0\over 2\pi} \left[k_0^2+\left(1-{\textstyle\frac{\lambda}{2}}\right)(2\ell+1)eB+\left({n\pi\over a}\right)^2+m^2\right]^{-s},
\label{zeta2}
\end{equation}
for mixed boundary conditions the zeta function is obtained by replacing $n$ in Eq. (\ref{zeta2}) with $n+{1\over 2}$.
When $u^\mu$ is spacelike and parallel to $\vec B$, the zeta function for Dirichlet boundary conditions is
\begin{equation}
\zeta(s)=\mu^{2s}{L^2eB\over 2\pi}\sum_{n=0}^\infty \sum_{\ell=0}^\infty\int^\infty_{-\infty} {dk_0\over 2\pi} \left[k_0^2+(2\ell+1)eB+(1-\lambda)\left({n\pi\over a}\right)^2+m^2\right]^{-s},
\label{zeta3}
\end{equation}
and the zeta function for mixed boundary conditions has $n$ replaced with $n+{1\over 2}$.

In the next three sections I will evaluate the zeta function and obtain the Casimir energy for Dirichlet and mixed boundary conditions in the three cases of timelike Lorentz anisotropy, spacelike anisotropy in the $x$-$y$ plane, and spacelike anisotropy in the $z$ direction.
\section{Timelike anisotropy}
\label{3}
I begin by investigating Dirichlet boundary conditions. I change the variable of integration in the $k_0$-integral of Eq. (\ref{zeta1}), and use the identity
\begin{equation}
z^{-s}={1\over \Gamma(s)}\int_0^\infty  t^{s-1}e^{-zt} dt,
\label{z}
\end{equation}
where $\Gamma(s)$ is the Euler gamma function, to obtain
\begin{equation}
\zeta(s)={\mu^{2s}\over \Gamma(s)}{L^2eB\over 4\pi^2}{ \sqrt{1+\lambda}}\sum_{n=0}^\infty \sum_{\ell=0}^\infty\int^\infty_{-\infty} \!\!\!\!dk_0 \int_0^\infty \!\!\!\! t^{s-1}e^{-\left[k_0^2+(2\ell+1)eB+({n\pi\over a})^2+m^2\right]t} dt\, .
\label{zetat1}
\end{equation}
I do the $k_0$ integration and use
\begin{equation}
\sum_{\ell=0}^\infty e^{-(2\ell +1)eBt}={1\over 2 \sinh(eBt)},
\label{sinh}
\end{equation}
to find
\begin{equation}
\zeta(s)={\mu^{2s}\over \Gamma(s)}{L^2eB\over 8\pi^{3\over 2}}\sqrt{1+\lambda}\sum_{n=0}^\infty  \int_0^\infty \!\!\!\! t^{s-3/2}e^{-\left[({n\pi\over a})^2+m^2\right]t}\csch{(eBt})dt .
\label{zetat2}
\end{equation}
This zeta function cannot be evaluated in a closed form in the general case, but can be reduced to a very simple form in some limiting cases. Notice that the term with $n=0$ is independent of $a$, does not contribute to the measurable Casimir pressure and therefore can be neglected.

In the short plate distance limit, $a^{-1}\gg \sqrt{eB}$ and  $a^{-1}\gg m$,
I take $e^{-m^2t}\simeq 1-m^2t$ and 
\begin{equation}
eB\csch(eBt) \simeq {1\over t}-{e^2B^2\over 6}t,
\label{sinh2}
\end{equation}
to obtain
\begin{equation}
\zeta(s)={\mu^{2s}\over \Gamma(s)}{L^2\over 8\pi^{3\over 2}}\sqrt{1+\lambda}\sum_{n=1}^\infty  \int_0^\infty \!\!\!\! t^{s-5/2}e^{-{n^2\pi^2\over a^2}t}\left(1-{e^2B^2\over 6}t^2\right)(1-m^2t)dt .
\label{zetat3}
\end{equation}
At this point, it is straightforward to write the zeta function for mixed boundary conditions by replacing $n$ with $n+{1\over 2}$ in Eq. (\ref{zetat3}), and retaining the $n=0$ term in the sum, since it is not independent of $a$,
\begin{equation}
\zeta(s)={\mu^{2s}\over \Gamma(s)}{L^2\over 8\pi^{3\over 2}}\sqrt{1+\lambda}\sum_{n=0}^\infty  \int_0^\infty \!\!\!\! t^{s-5/2}e^{-{\left(n+{1/2}\right)^2\pi^2\over a^2}t}\left(1-{e^2B^2\over 6}t^2\right)(1-m^2t)dt .
\label{zetat3m}
\end{equation}

After I do the integration and neglect a smaller term proportional to $e^2B^2m^2$ I find, for Dirichlet boundary conditions
\begin{eqnarray}
\zeta(s)=&&{\mu^{2s}\over \Gamma(s)}{L^2\over 8\pi^{3\over 2}}{1\over \sqrt{1-\lambda}}\left({\pi\over a}\right)^{3-2s}\biggl[\Gamma(s-{\textstyle\frac{3}{2}})\zeta_R(2s-3)-{m^2a^2\over \pi^2}\Gamma(s-{\textstyle\frac{1}{2}})\zeta_R(2s-1)
\nonumber \\
&&-\,\,{e^2B^2a^4\over 6\pi^4}\Gamma(s+{\textstyle\frac{1}{2}})\zeta_R(2s+1)\biggr],
\label{zetat4}
\end{eqnarray}
where $\zeta_R(s)$ is the Riemann zeta function. For mixed boundary conditions I find
\begin{eqnarray}
\zeta(s)=&&{\mu^{2s}\over \Gamma(s)}{L^2\over 8\pi^{3\over 2}}\sqrt{1+\lambda}
\left({\pi\over a}\right)^{3-2s}\biggl[\Gamma(s-{\textstyle\frac{3}{2}})\zeta_H(2s-3,{\textstyle\frac{1}{2}})-{m^2a^2\over \pi^2}\Gamma(s-{\textstyle\frac{1}{2}})\zeta_H(2s-1,{\textstyle\frac{1}{2}})
\nonumber \\
&&-\,\,{e^2B^2a^4\over 6\pi^4}\Gamma(s+{\textstyle\frac{1}{2}})\zeta_H(2s+1,{\textstyle\frac{1}{2}})\biggr],
\label{zetat4m}
\end{eqnarray}
where $\zeta_H(s,x)$ is the Hurwitz zeta function defined as
\begin{equation}
\zeta_H(s,x) = \sum_{n=0}^\infty (n+x)^{-s}.
\label{Hurwitz}
\end{equation}

 In the case of a complex scalar field the Casimir energy, $E_C$, is related to the derivative of $\zeta(s)$ by
\begin{equation}
E_C = -\zeta'(0).
\label{E_C}
\end{equation}
Using the following power series expansions, valid for $s\ll 1$,
\begin{equation}
\left({\mu a\over \pi}\right)^{2s}{\Gamma(s-{\textstyle\frac{3}{2}})\over\Gamma(s)}\zeta_R(2s-3)\simeq {\sqrt{\pi}\over 90}s +{\cal O}(s^2),
\label{id1}
\end{equation}
\begin{equation}
\left({\mu a\over \pi}\right)^{2s}{\Gamma(s-{\textstyle\frac{3}{2}})\over\Gamma(s)}\zeta_H(2s-3,{\textstyle\frac{1}{2}})\simeq -{7\sqrt{\pi}\over 720}s +{\cal O}(s^2),
\label{id1m}
\end{equation}
\begin{equation}
\left({\mu a\over \pi}\right)^{2s}{\Gamma(s-{\textstyle\frac{1}{2}})\over\Gamma(s)}\zeta_R(2s-1)\simeq {\sqrt{\pi}\over 6}s +{\cal O}(s^2),
\label{id2}
\end{equation}
\begin{equation}
\left({\mu a\over \pi}\right)^{2s}{\Gamma(s-{\textstyle\frac{1}{2}})\over\Gamma(s)}\zeta_H(2s-1,{\textstyle\frac{1}{2}})\simeq -{\sqrt{\pi}\over 12}s +{\cal O}(s^2),
\label{id2m}
\end{equation}
\begin{equation}
\left({\mu a\over \pi}\right)^{2s}{\Gamma(s+{\textstyle\frac{1}{2}})\over\Gamma(s)}\zeta_R(2s+1)\simeq {\sqrt{\pi}\over 2}+{\sqrt{\pi}}\left[\gamma_E+\ln\left({\mu a\over 2\pi}\right)\right]s +{\cal O}(s^2),
\label{id3}
\end{equation}
\begin{equation}
\left({\mu a\over \pi}\right)^{2s}{\Gamma(s+{\textstyle\frac{1}{2}})\over\Gamma(s)}\zeta_H(2s+1,{\textstyle\frac{1}{2}})\simeq {\sqrt{\pi}\over 2}+{\sqrt{\pi}}\left[\gamma_E+\ln\left({2\mu a\over \pi}\right)\right]s +{\cal O}(s^2),
\label{id3m}
\end{equation}
where $\gamma_E=0.57721\cdots$ is the Euler-Mascheroni constant, I obtain the Casimir energy in the short plate distance limit for Dirichlet boundary conditions
\begin{equation}
{E_C} = -{\pi^2\over 8}{L^2\over a^3}{(1+\lambda)^{{1\over 2}}}\left({1\over 90}-{m^2a^2\over 6\pi^2}-{e^2B^2a^4\over 6\pi^4}\left[\gamma_E+\ln\left({\mu a\over 2\pi}\right)\right]\right),
\label{E_C1}
\end{equation}
and for mixed boundary conditions
\begin{equation}
{E_C} = {\pi^2\over 8}{L^2\over a^3}{(1+\lambda)^{{1\over 2}}}\left({7\over 720}-{m^2a^2\over 12\pi^2}+{e^2B^2a^4\over 6\pi^4}\left[\gamma_E+\ln\left({2\mu a\over \pi}\right)\right]\right),
\label{E_C1m}
\end{equation}
where I need to fix the parameter $\mu$ and will make the choice $\mu =\max\{m, \sqrt{eB}\}$. The arbitrary parameter $\mu$ could be fixed by additional conditions, if available, and, to some extent, its presence signals an ambiguity in the method. However, in most problems where the solution is obtained using the  zeta function technique, $\mu$ turns out to be the largest mass scale present in the problem.
The leading order terms of Eqs. (\ref{E_C1}) and (\ref{E_C1m}) agree with the leading terms calculated by Cruz et al. \cite{Cruz:2017kfo} in the short plate distance limit for Dirichlet and mixed boundary conditions respectively. However, these authors \cite{Cruz:2017kfo} miss the leading order correction of order $(ma)^2$ that I obtain here, in spite of their general procedure being correct. The $ma \ll 1$ asymptotic limit of their result, which is expressed as an integral without analytic solution, has a lower degree of precision than the calculation presented here, based on a different method, which represents the next step of the study. Notice that my results, when compared to those of Ref. \cite{Cruz:2017kfo}, carry an extra factor of two because I am considering a complex scalar field, not a real one. Finally, Eqs. (\ref{E_C1}) and (\ref{E_C1m}) show that the timelike anisotropy causes a slight increase of the magnetic field correction to $E_C$, when compared to the magnetic correction to $E_C$ in isotropic spacetime.

In the strong magnetic field limit, $\sqrt{eB}\gg a^{-1} , m$. In the case of Dirichlet boundary conditions, I start with Eq. (\ref{zetat2}), approximate $\csch(eBt)\simeq 2e^{-eBt}$ and do a Poisson resummation of the series to find
\begin{equation}
\zeta(s)={\mu^{2s}\over \Gamma(s)}{eBL^2a\over 4\pi^{2}}\sqrt{1+\lambda}\sum_{n=1}^\infty  \int_0^\infty \!\!\!\! t^{s-2}e^{-{n^2a^2\over t}}e^{-(eB+m^2)t} dt,
\label{zetat5}
\end{equation}
where, again, I neglected the $n=0$ term which is easily evaluated to be
\begin{equation}
\zeta_0(s)={L^2a\over 8\pi^{2}}{(eB+m^2)eB}\sqrt{1+\lambda}\left({\mu^2\over eB+m^2}\right)^s{\Gamma(s-1)\over \Gamma(s)},
\label{zeta0}
\end{equation}
and produces a uniform energy density term that does not contribute to the Casimir pressure as long as the magnetic field is present inside and outside the plates \cite{Erdas:2013dha}. 
I change integration variable in Eq. (\ref{zetat5}) and find
\begin{equation}
\zeta(s)={\mu^{2s}\over \Gamma(s)}{eBL^2a\over 4\pi^{2}}\sqrt{1+\lambda}\left({a\over \sqrt{eB+m^2}}\right)^{s-1}\sum_{n=1}^\infty  \int_0^\infty \!\!\!\! t^{s-2}e^{-na\sqrt{eB+m^2}(t+1/t)} dt.
\label{zetat6}
\end{equation}
For mixed boundary conditions I follow similar steps and find
\begin{equation}
\zeta(s)={\mu^{2s}\over \Gamma(s)}{eBL^2a\over 4\pi^{2}}\sqrt{1+\lambda}\left({a\over \sqrt{eB+m^2}}\right)^{s-1}\sum_{n=1}^\infty  (-1)^n\int_0^\infty \!\!\!\! t^{s-2}e^{-na\sqrt{eB+m^2}(t+1/t)} dt.
\label{zetat6m}
\end{equation}

In the strong magnetic field limit $a\sqrt{eB}\gg 1$, and all terms with $n>1$ in the sums of Eqs. (\ref{zetat6}) and (\ref{zetat6m}) are negligibly small and can be left out. I evaluate the remaining integrals using the saddle point method and find, for Dirichlet boundary conditions
\begin{equation}
\zeta(s)=\left({\mu^{2}a\over \sqrt{eB+m^2}}\right)^s{L^2eB\over 4\pi^{3/2}}\sqrt{1+\lambda}{(eB+m^2)^{1\over 4}\over \sqrt{a}} {e^{-2a\sqrt{eB+m^2}}\over \Gamma(s)}.
\label{zetat7}
\end{equation}
The derivative of $\zeta(s)$ is easily calculated using 
\begin{equation}
{A^s\over \Gamma(s)}\simeq s+{\cal O}(s^2), 
\label{id4}
\end{equation}
to obtain the strong magnetic field Casimir energy for Dirichlet boundary conditions
\begin{equation}
E_C=-{L^2eB\over 4\pi^{3/2}}\sqrt{1+\lambda}{(eB+m^2)^{1\over 4}\over \sqrt{a}} e^{-2a\sqrt{eB+m^2}}.
\label{E_C2}
\end{equation}
and, with the same method, for mixed boundary conditions
\begin{equation}
E_C={L^2eB\over 4\pi^{3/2}}\sqrt{1+\lambda}{(eB+m^2)^{1\over 4}\over \sqrt{a}} e^{-2a\sqrt{eB+m^2}}.
\label{E_C2m}
\end{equation}
Notice that the dominant term is, in both cases, the exponential suppression term.

Finally I will examine the large mass limit, $m\gg \sqrt{eB}, a^{-1}$. I do a Poisson resummation of the zeta function of Eq. (\ref{zetat2}) and find
\begin{equation}
\zeta(s)={\mu^{2s}\over \Gamma(s)}{L^2aeB\over 8\pi^{2}}{\sqrt{1+\lambda}}\sum_{n=1}^\infty  \int_0^\infty \!\!\!\! t^{s-2}e^{-\left(m^2t+{n^2a^2\over t}\right)}\csch{(eBt})dt ,
\label{zetat8}
\end{equation}
after neglecting the $n=0$ term in the sum because it is a uniform energy density term that will not contribute to the Casimir energy. Next, I change the integration variable to find
\begin{equation}
\zeta(s)=\left({\mu^{2}a\over m}\right)^s{eBL^2m\over 8\pi^{2}\Gamma(s)}{\sqrt{1+\lambda}}\sum_{n=1}^\infty  \int_0^\infty \!\!\!\! t^{s-2}e^{-nam\left(t+{1\over t}\right)}\csch{(neBat/m)}dt ,
\label{zetat9}
\end{equation}
where all the terms with $n>1$ are negligible because $am\gg1$. I do the integral using the saddle point method, use (\ref{id4}) to take the derivative of $\zeta (s)$, and find the large mass limit of the Casimir energy for Dirichlet boundary conditions
\begin{equation}
E_C=-{L^2\over 8}{\sqrt{1+\lambda}}\left({m\over \pi a}\right)^{3\over2}{e^{-2am}}F(z) , 
\label{E_C3}
\end{equation}
where $z=eBa/m$ is dimensionless and 
\begin{equation}
F(z)=z\csch(z)
\label{Fz}
\end{equation}
is the magnetic field correction to $E_C$. The large mass limit for mixed boundary conditions is found by taking similar steps, and is given by Eq. (\ref{E_C3}) without the initial negative sign.
Notice that $F(z)\rightarrow 1$ when $z\rightarrow 0$, and thus my result agrees with the large mass limit of Ref. \cite{Cruz:2017kfo} when $B\rightarrow 0$.
\section{Spacelike anisotropy in the x-y plane}
\label{4}
When the Lorentz anisotropy is spacelike and perpendicular to $\vec B$, the zeta function (\ref{zeta2}) for Dirichlet boundary conditions can be written as
\begin{equation}
\zeta(s)={\mu^{2s}\over \Gamma(s)}{L^2eB\over 4\pi^2}\sum_{n=0}^\infty \sum_{\ell=0}^\infty\int^\infty_{-\infty} \!\!\!\!dk_0 \int_0^\infty \!\!\!\!dt t^{s-1}e^{-\left[k_0^2+(1-{\lambda\over 2})(2\ell+1)eB+({n\pi\over a})^2+m^2\right]t},
\label{zetas1}
\end{equation}
and the zeta function for mixed boundary conditions is obtained from it by replacing $n$ with $n+{1\over 2}$.
Comparing (\ref{zetas1}) with (\ref{zetat1}), it is evident that (\ref{zetas1}) can be obtained by taking (\ref{zetat1}) and replacing $eB$ with $(1-{\lambda\over 2})eB$ in it. The same replacement works when obtaining the zeta function for mixed boundary conditions from its timelike asymmetry analogous. Therefore $E_C$ for anisotropy in the $x-y$ plane is obtained, in each of the three asymptotic limits for Dirichlet and mixed boundary conditions, by taking Eqs. (\ref{E_C1}), (\ref{E_C2}), (\ref{E_C3}) and making the same replacement.

In the short plate distance limit, I find
\begin{equation}
{E_C} = -{\pi^2\over 8}{L^2\over a^3}{(1-\lambda)^{-{1\over 2}}}\left({1\over 90}-{m^2a^2\over 6\pi^2}-{(1-\lambda)e^2B^2a^4\over 6\pi^4}\left[\gamma_E+\ln\left({\mu a\over 2\pi}\right)\right]\right),
\label{E_C4}
\end{equation}
for Dirichlet boundary conditions, and
\begin{equation}
{E_C} = {\pi^2\over 8}{L^2\over a^3}{(1-\lambda)^{-{1\over 2}}}\left({7\over 720}-{m^2a^2\over 12\pi^2}+{(1-\lambda)e^2B^2a^4\over 6\pi^4}\left[\gamma_E+\ln\left({2\mu a\over \pi}\right)\right]\right),
\label{E_C4m}
\end{equation}
for mixed boundary conditions. Eqs. (\ref{E_C4}) and  (\ref{E_C4m})
show a slight decrease of the magnetic field contribution to $E_C$ when compared to the cases of timelike anisotropy and isotropic spacetime. As I did in Eqs.  (\ref{E_C1}) and (\ref{E_C1m}), I choose $\mu=\max\{m,\sqrt{eB}\}$. When $B\rightarrow 0$, the leading order term of these results agrees with the leading order terms obtained in Ref. \cite{Cruz:2017kfo} under the same conditions.

In the strong magnetic field limit, the Casimir energy for Dirichlet boundary conditions is given by
\begin{equation}
E_C=-{L^2eB\over 4\pi^{3/2}}{\left[(1-{\lambda\over 2})eB+m^2\right]^{1\over 4}\over \sqrt{a}} e^{-2a\sqrt{(1-{\lambda\over 2})eB+m^2}}.
\label{E_C5}
\end{equation}
and the Casimir energy for mixed boundary conditions is obtained from (\ref{E_C5}) by removing the initial negative sign.
Finally, in the large mass limit, I obtain for Dirichlet boundary conditions
\begin{equation}
E_C=-{L^2\over 8}\left({m\over \pi a}\right)^{3\over 2}{e^{-2am}\over \sqrt{1-\lambda}}F(z\sqrt{1-\lambda}) , 
\label{E_C6}
\end{equation}
and for mixed boundary conditions I obtain the same without the initial negative sign. These results are
in agreement with the large mass limit of Ref. \cite{Cruz:2017kfo} when $B\rightarrow 0$. Notice that the magnetic correction obtained here is smaller than the one obtained in the case of timelike anisotropy.
\section{Spacelike anisotropy in the z direction}
\label{5}
When the spacelike Lorentz anisotropy is parallel to $\vec B$ and Dirichlet boundary conditions are assumed, the zeta function can be written as
\begin{equation}
\zeta(s)={\mu^{2s}\over \Gamma(s)}{eBL^2\over 4\pi^2}\sum_{n=0}^\infty \sum_{\ell=0}^\infty\int^\infty_{-\infty} \!\!\!\!dk_0 \int_0^\infty \!\!\!\!dt t^{s-1}e^{-\left[k_0^2+(2\ell+1)eB+(1-{\lambda})({n\pi\over a})^2+m^2\right]t}.
\label{zetas2}
\end{equation}
This zeta function can be obtained immediately by multiplying the zeta function of Eq. (\ref{zetat1}) by $\sqrt {1-\lambda}$ and then replacing $a$ with $a\over \sqrt {1-\lambda}$ in it. The same replacement, applied to the zeta function for timelike asymmetry and mixed boundary conditions, produces the zeta function for mixed boundary conditions.Therefore $E_C$ for anisotropy in the $z$ direction is  obtained  by making the same modifications to the three asymptotic forms of $E_C$ of Eqs. (\ref{E_C1}), (\ref{E_C1m}),  (\ref{E_C2}), (\ref{E_C2m}), and (\ref{E_C3}).

In the limit of short plate distance and Dirichlet boundary conditions, I obtain
\begin{equation}
{E_C} = -{\pi^2\over 8}{L^2\over a^3}{(1-\lambda)^{{3\over 2}}}\left[{1\over 90}-{m^2a^2\over 6\pi^2(1-\lambda)}-{e^2B^2a^4\over 6\pi^4(1-\lambda)^2}\left(\gamma_E+\ln\left({\mu a\over 2\pi}\right)+{\lambda\over 2}\right)\right],
\label{E_C7}
\end{equation}
and for mixed boundary conditions
\begin{equation}
{E_C} = {\pi^2\over 8}{L^2\over a^3}{(1-\lambda)^{{3\over 2}}}\left[{7\over 720}-{m^2a^2\over 12\pi^2(1-\lambda)}+{e^2B^2a^4\over 6\pi^4(1-\lambda)^2}\left(\gamma_E+\ln\left({2\mu a\over \pi}\right)+{\lambda\over 2}\right)\right],
\label{E_C7m}
\end{equation}
where I used $\ln[(1-\lambda)^{-{1\over 2}}] \simeq \lambda/2$ and $\mu=\max\{m,\sqrt{eB}\}$. When $B\rightarrow 0$, the leading order terms of the last two equations agree with the leading order terms of the short plate distance limit of Ref. \cite{Cruz:2017kfo} for anisotropy in the $z$ direction. Notice that the magnetic field correction, apart from the smaller term proportional to $\lambda/2$, is the same in this limit as it is for a timelike anisotropy when the plate distance is short, as shown in Eqs. (\ref{E_C1}) and (\ref{E_C1m}).

In the strong magnetic field limit I find the following Casimir energy for Dirichlet boundary conditions
\begin{equation}
E_C=-{L^2eB\over 4\pi^{3\over 2}}{\left(eB+m^2\right)^{1\over 4}\over \sqrt{a}} (1-{\lambda})^{1\over 4}e^{-2{a}\sqrt{eB+m^2\over 1-\lambda}},
\label{E_C8}
\end{equation}
while the Casimir energy for mixed boundary conditions is obtained by eliminating the initial negative sign in Eq. (\ref{E_C8}).
Last, in the large mass limit and with Dirichlet boundary conditions, I obtain
\begin{equation}
E_C=-{L^2\over 8}\left({m\sqrt{1-\lambda}\over \pi a}\right)^{3\over 2}e^{-2{am\over \sqrt{1-\lambda}} }F(z/\sqrt{1-\lambda}) , 
\label{E_C9}
\end{equation}
and for mixed boundary conditions $E_C$ is given by the last result multiplied by negative one. These results, when $B\rightarrow 0$, agree with those obtained in Ref. \cite{Cruz:2017kfo}.
\section{Casimir pressure}
\label{6}
The Casimir pressure is given by
\begin{equation}
P_C=-{1\over L^2}{\partial E_C\over \partial a}.
\label{P_C1}
\end{equation}
Keeping the assumption that the magnetic field is present inside and outside the plates, all uniform energy density terms neglected previously will not contribute to the Casimir pressure. Furthermore, once the Casimir pressure for a timelike anisotropy is presented, the pressure in the cases of spacelike anisotropy in the $z$ direction or $x-y$ plane is immediately obtained with a simple substitution. In the presence of a timelike anisotropy, the short plate distance Casimir pressure for Dirichlet boundary conditions is
\begin{equation}
{P_C} = -{\pi^2\over 8a^4}{(1+\lambda)^{{1\over 2}}}\left({1\over 30}-{m^2a^2\over 6\pi^2}+{e^2B^2a^4\over 6\pi^4}\left[1+\gamma_E+\ln\left({\mu a\over 2\pi}\right)\right]\right),
\label{P_C2}
\end{equation}
and for mixed boundary conditions
\begin{equation}
{P_C} = {\pi^2\over 8a^4}{(1+\lambda)^{{1\over 2}}}\left({7\over 240}-{m^2a^2\over 12\pi^2}-{e^2B^2a^4\over 6\pi^4}\left[1+\gamma_E+\ln\left({2\mu a\over \pi}\right)\right]\right),
\label{P_C2m}
\end{equation}
where the parameter $\mu$ is $\mu =\max\{m, \sqrt{eB}\}$. Notice the usual Casimir effect feature, where pressure is attractive for Dirichlet boundary conditions and repulsive for mixed boundary conditions. In both cases, the presence of a magnetic field contributes an almost constant attractive term, aside from a weak logarithmic dependence on $a$. The short plate distance pressure in the presence of 
a spacelike anisotropy in the $x-y$ direction is obtained from Eqs. (\ref{P_C2}) and (\ref{P_C2m}) by replacing $eB$ with $(1-{\lambda\over 2})eB$, and the short plate distance pressure for anisotropy in the $z$ direction  is obtained by replacing $a$ with $a\over \sqrt {1-\lambda}$ in Eqs. (\ref{P_C2}) and (\ref{P_C2m}).

The strong magnetic field limit of the Casimir pressure, in the presence of a timelike anisotropy and Dirichlet boundary conditions, is
\begin{equation}
P_C=-{eB\over 2\pi^{3/2}}{(eB+m^2)^{3\over 4}\over {\sqrt a}} (1-\lambda)^{-{1\over 2}}e^{-2a\sqrt{eB+m^2}}\left(1+{1\over 4a\sqrt{eB+m^2}}\right),
\label{P_C3}
\end{equation}
while the Casimir pressure for mixed boundary conditions does not carry the initial negative sign.
Here the dominant term is the exponential suppression, since $a\sqrt{eB}\gg 1$. In the case of a spacelike anisotropy in the $x-y$ direction, the strong field limit of the pressure is obtained from (\ref{P_C3}) by replacing $eB$ with $(1-{\lambda\over 2})eB$, which produces a slightly weaker exponential suppression of $P_C$, when compared to timelike anisotropy or isotropic spacetime. On the other hand, the pressure in the strong field limit for anisotropy in the $z$ direction has a slightly increased exponential suppression when compared to a timelike anisotropic or isotropic spacetime. This expression of $P_C$ is obtained from (\ref{P_C3}) by replacing $a$ with $a\over \sqrt {1-\lambda}$.

Last, the Casimir pressure in the large mass limit in the presence of a timelike anisotropy and Dirichlet boundary conditions, is
\begin{equation}
P_C=-{m\over 4}\left({m\over \pi a}\right)^{3/2}{e^{-2am}\over \sqrt{1-\lambda}}F(z) \left(1+{1\over 4am}+{eB\over 2m^2}\coth z\right), 
\label{P_C4}
\end{equation}
where $z={eBa\over m}$ and $F(z)$ is defined in (\ref{Fz}). The large mass limit for timelike anisotropy and mixed boundary conditions is the same, but without the initial $-1$ factor. Notice that, also here, the exponential suppression is the dominant term, since $am\gg1$. The large mass limit of the pressure in the case of spacelike anisotropy, either in the $z$ or in the $x-y$ direction, is obtained from (\ref{P_C4}) by making the substitutions outlined in the previous paragraph.
\section{Discussion and conclusions}
\label{7}
In this work I used the generalized zeta function technique to study the Casimir effect of a Lorentz-violating scalar field in the presence of a magnetic field. This massive and charged complex scalar field satisfies a modified Klein-Gordon equation that breaks Lorentz symmetry in a CPT-even aether-like manner by the direct coupling of the field derivative to a constant unit four-vector $u^\mu$. I investigated the case of this field having Dirichlet boundary conditions, and mixed (Dirichlet-Neumann) boundary conditions on two plane parallel plates, perpendicular to the magnetic field. All the results obtained for Dirichlet boundary conditions are valid also in the case of the field satisfying Neumann boundary conditions on the plates. I obtained simple analytic expressions for the Casimir energy in the asymptotic cases of short plate distance, strong magnetic field and large mass when $u^\mu$ is timelike, Eqs. (\ref{E_C1}) and (\ref{E_C1m}), (\ref{E_C2}) and (\ref{E_C2m}), and (\ref{E_C3}) respectively; when  $u^\mu$ is spacelike and orthogonal to $\vec B$, Eqs. (\ref{E_C4}) and (\ref{E_C4m}), (\ref{E_C5}), (\ref{E_C6}); when  $u^\mu$ is spacelike and parallel to $\vec B$, Eqs. (\ref{E_C7}) and (\ref{E_C7m}), (\ref{E_C8}), (\ref{E_C9}).
In Sec. \ref{6} I obtained simple analytic expressions of the Casimir pressure in the case of $u^\mu$ timelike, spacelike and orthogonal to $\vec B$ and spacelike and parallel to $\vec B$ in the three asymptotic cases. 

The paper by Cruz et al.\cite{Cruz:2017kfo} presents numerical solutions for their results
of the Casimir effect in Lorentz-violating scalar field theory and they examine values of the Lorentz-violating parameter $\lambda = 0, 0.1, 0.2$. In the case of strong magnetic field, the correction to the Casimir energy that I obtain for a value of $\lambda = 0.1$, when compared to the case of the Lorentz-symmetric case with $\lambda =0$, amounts to $10\% - 15\%$, depending on boundary conditions and direction of the Lorentz symmetry violation. In the case of short plate distance and large mass, I find corrections to $E_C$ due to magnetic field in the range of $1.0\% - 1.5\%$, when compared to the case where $\lambda  = 0$. The corrections in the case of $\lambda = 0.2$ are, roughly, doubled.

However, such estimates of $\lambda$ are, most likely, too high. Many papers have been published that examine Lorentz violation within QED, through the Casimir effect \cite{Frank:2006ww,Kharlanov:2009pv}, or other means \cite{Klinkhamer:2010zs,Exirifard:2010xm,Casana:2011vh,Casana:2011du}. Ref. \cite{Kharlanov:2009pv}, for example, found that, within the framework of the (3+1)-dimensional Lorentz-violating extended electrodynamics, the leading $\lambda$-correction to the Casimir force is only about $1\% - 2\%$ of its value for $\lambda = 0$. An earlier paper \cite{Frank:2006ww} analyzed extended QED from
the minimal version of the Standard Model including Lorentz-violating terms for both the scalar and fermion sectors.
They reported an additive correction to the Casimir force due to CPT-odd Lorentz-violating terms in the Lagrangian that has a $1/a^2$ dependence on the plates distance, and is consistent within $15\%$ with the experimental data. Other papers looking at direct experiments not involving the Casimir effect, or looking at indirect bounds of Lorentz violation, place Lorentz-violating effects within QED to a part in $10^{-8}$, or less. These papers look at a variety of factors to obtain these bounds, from birefringent constraints on light \cite{Exirifard:2010xm}, to observation of ultra-high-energy-cosmic-ray primaries and TeV gamma-rays at the top of the Earth's atmosphere \cite{Klinkhamer:2010zs}.
While these estimates are for QED, it seems reasonable that, for the scalar theory I am considering, I take the value of $\lambda\le 0.01 - 0.02$. In this case, the magnetic corrections I obtain and report in the previous paragraph, should be divided by ten and the effects being investigated are likely to be beyond experimental reach.

Finally, my results confirm the attractive Casimir force for Dirichlet and Neumann boundary conditions, and the repulsive force for mixed boundary conditions. The results obtained show a strong dependence of $E_C$ and $P_C$ on the dimensionless quantity $\lambda$ that parametrizes the breaking of the Lorentz symmetry. This strong dependence on $\lambda$ was also observed in Ref. \cite{Cruz:2017kfo}. When the magnetic field is switched off my results agree, to leading order, with the results of Ref. \cite{Cruz:2017kfo}, and also agree with the well known results for isotropic spacetime when $\lambda \rightarrow 0$. In the short plate distance limit, I find that the magnetic correction decreases the Casimir pressure in all three cases of a timelike and spacelike $u^\mu$, with the biggest correction happening  when $u^\mu$ is 
spacelike and parallel to $\vec B$. In the case of a strong magnetic field I discover that $P_C$ is exponentially suppressed in all three cases, and find that the pressure is largest when $u^\mu$ is spacelike and orthogonal to $\vec B$. Finally, I find that $P_C$ is exponentially suppressed also in the large mass limit, and the largest magnetic correction happens when $u^\mu$ is spacelike and parallel to $\vec B$.

\end{document}